%% file: paperKmecSchmitt.tex
\documentclass{svproc}

\usepackage{url}

\usepackage{amsmath}
\usepackage{graphicx,xcolor,enumerate}
\ifx\comment\undefined
\newcommand{\comment}[1]{}
\else
\fi

\newcommand{\m}[1]{\ensuremath{\mathrm{#1}}}
\newcommand{\NHH}{\ensuremath{N_\m{HH/section}}}
\begin{document}
\mainmatter              % start of a contribution
\title{Exploring the fitness landscape of a realistic turbofan rotor blade optimization }
\titlerunning{\texttt{https://doi.org/10.1007/978-3-319-97773-7\_46} }
\author{Jakub Kmec\inst{1} \and Sebastian Schmitt\inst{2}}
\authorrunning{J.\ Kmec and S.\ Schmitt} % abbreviated author list (for running head)
\tocauthor{Jakub Kmec\inst{1}, Sebastian Schmitt}
\institute{Palack\'y University,\\ Olomouc, Czech Republic\\
\email{jakub.kmec01@upol.cz}
\and
Honda Research Institute Europe GmbH,\\
Offenbach, Germany\\
\email{sebastian.schmitt@honda-ri.de}
}

\maketitle              % typeset the title of the contribution

\begin{abstract}
 \input{abstract.tex}
\end{abstract}
\small{
  Published as: Kmec J., Schmitt S. (2019) Exploring the Fitness Landscape of a Realistic Turbofan Rotor Blade Optimization. In: Rodrigues H. et al. (eds) EngOpt 2018 Proceedings of the 6th International Conference on Engineering Optimization. EngOpt 2018.\\
https://doi.org/10.1007/978-3-319-97773-7\_46
}

\section{Introduction}

Aerodynamic shape optimization using computational fluid dynamics (CFD)
is widely used  for creating
efficient designs in engineering applications such as the automotive or the aerospace domain \cite{designOpt}.
Global optimization techniques enable unbiased search for improved shapes and are capable of
yielding conceptional new designs. However, they usually require a significant number of CFD simulations
for assessing the  quality of the proposed designs. This constitutes a substantial 
drawback for extensive application of global optimization techniques in actual design cycles since realistic
CFD simulations are usually rather complex and have a large time and resource demand.
Difficulties arise in that context due to the many options available to the designer for setting
up such a numerical design optimization.
These include the choice of the optimizer algorithm, specific configuration of the parameters
for the optimizer algorithm, representation of the design and design changes to be used, and many more. 

In practice, one usually employs a specific setup with some more or less well motivated choices. 
Then optimization is performed and depending on the results some refinements are done and a new optimization run is performed. 
This is iterated until a satisfactory result is obtained.  
It is not possible to assess whether the results obtained in this manner are representative. 
In particular, it is not clear if the results could have been much better,
if a slightly different configuration was chosen. Or, if a rather different design
could have been obtained which still has a similarly good fitness value.
The answers to such questions are of course strongly linked to the qualitative behavior of
the fitness function.

In this work, we explore the fitness landscape of a realistic turbofan 
simulation case. We investigate the outcome of different optimization runs in which we vary
the configuration settings. We compare the achievable fitness values obtained with
a covariance matrix adaptation evolutionary strategy (CMA-ES) \cite{cma} to a
particle swarm optimization (PSO) \cite{pso}.

Another  key ingredient to shape optimization
approaches is the choice how to represent a specific design and in particular how to generate
sets of different designs to be evaluated during the optimization. In a typical engineering
application, a valid baseline design is readily available and the design process focuses on
improving that design. Deformation methods are especially suited and well-established
for such situations. A crucial decision to be taken by the engineer is the choice of the
deformation method and the details of its realization. In the context of evolutionary
aerodynamic shape optimization, shape morphing methods are popular such as free-form deformation (FFD)
and radial basis functions (RBF) based deformation methods, or shape-function methods like
Hicks-Henne (HH) functions \cite{deform}.

%-------------------------------------------------------------------------
\section{Optimization and simulation setup}

\subsection{Blade representation}

\begin{figure}[!htb]  
  \centering
  \includegraphics[height=0.27\linewidth]{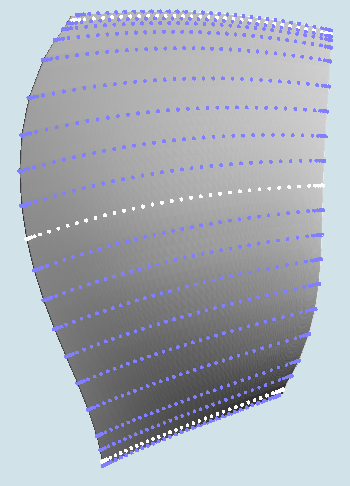} %pics/bladeSectionsSketch.png}
  \includegraphics[height=0.3\linewidth]{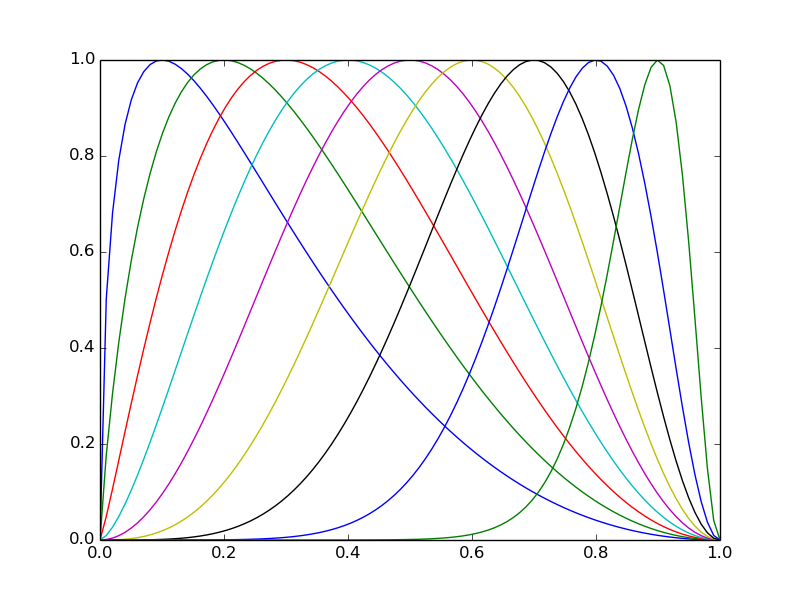}
  \caption{\label{HicksHenne}
    Left: 3D baseline blade
    geometry where the points for the sections are shown as blue dots and the three
    independently deformable sections near the hub, the mid-span and the shroud are shown in white.
    Right: Hicks-Henne (HH) shape functions. } 
\end{figure}

During the optimization, the blade geometry has to be changed automatically based on the parameters
proposed by the optimization algorithm. Thus, a blade representation has to be defined. In this
work, we use a deformation approach, where we start from a given baseline blade design and
encode changes to it via the optimization parameters. 
Specifically, a  blade is represented as a collection of sections stacked on top of each other as shown in
Figure \ref{HicksHenne} (left). Each section (which can be thought of as a 2D airfoil profile)
is defined by a cylindrical cut of the blade geometry
where the same number of sample points is used for each section. 
The 3D geometry is reconstructed from those sections by linearly interpolating between the sections.

In the most general situation, all sections would be allowed to change independently to generate a modified
blade design.
In order to keep the optimization parameters manageable, we  select only  three sections, one
close to the hub, one in the mid-span region, and one close to the shroud, which we
deform based on the optimization parameters. The in-between sections are linearly interpolated.

For each of those independently deformed sections, we allow the following changes: (i) rotation of the section
around the leading edge point, (ii) movement of the section in the axial-meridional plane, and (iii) deformation of the section
profile by adding HH shape functions (see Figure \ref{HicksHenne} right), which is a well-known
representation from 2D airfoil design, see e.g. \cite{HicksHenne}.%wuHH03}. 
The explicit functional form
of the HH shape  functions we use  is  
\begin{align}
b(x,x_0) = \Big[\sin\Big(\pi x^{\frac{\log(0.5)}{\log(x_0)}}\Big)   \Big]^2,
\end{align}
where $x \in [0,1]$ parametrizes the chord length of each section and $x_0$ controls the location
of the maximum of each shape function.
The number of shape functions per section \NHH will be varied in this work and for a given number  of 
$\NHH$ we use equally spaced
locations for the maxima, i.e.\ $x_0(i)=\frac{i}{\NHH+1}$ where $i=1,\dots,\NHH$.

In order to fully characterize the deformed blade geometry, we need to specify for each of the three independently 
deformable sections the amplitude for each shape function, the rotation angle, and the shift in axial and 
rotational direction. 
Therefore, the total number of free parameters, which is also the number of search 
variables for the optimization problem, is given by 
$
  \label{eqSearchDim}
  N_\m{search}=3(\NHH+3).
$.

\subsection{Simulation setup}

The turbo-fan simulation setup was inspired by the GE Honda HF120 \cite{hf120} small business jet engine
operated at cruise condition.
We used \texttt{steadyCompressibleMRFFoam}, the compressible flow solver from the OpenFOAM CFD
suite (specifically, the \texttt{foam-extend-3.2} version) which was adapted to be more
robust \cite{ruscheTalk16}.
The blade rotation speed was set to a fixed value of 15600rpm
and constant
mass-flow and radial equilibrium boundary conditions were employed. We used the $k\omega$SST turbulence model. 
The meshes for all simulations consisted of about 6 million cells and were created by an
in-house software based 
on the OpenFOAM utility \texttt{blockMesh}. 
The flow equations were solved on 32 CPU cores in parallel, iterated for 12000 solver iterations
in order to arrive at a converged solution, and the run-time was typically in the range of 2-4 hours. 

\begin{figure}[htb!]
  \centering
  \includegraphics[height=0.3\linewidth]{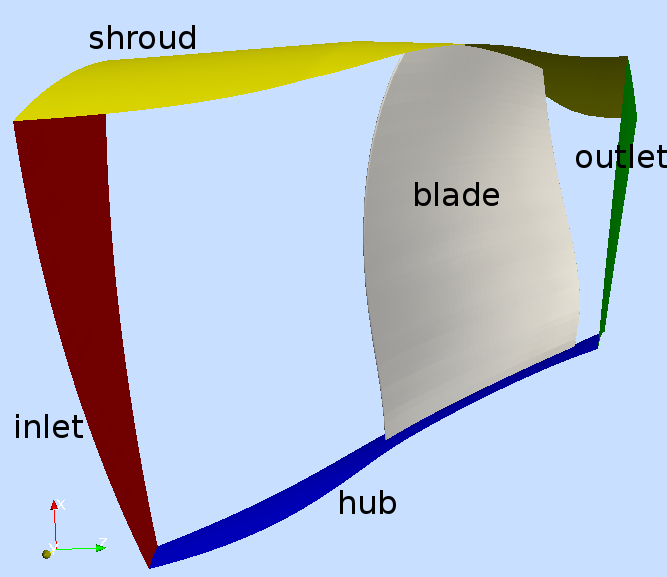}
  \includegraphics[height=0.3\linewidth]{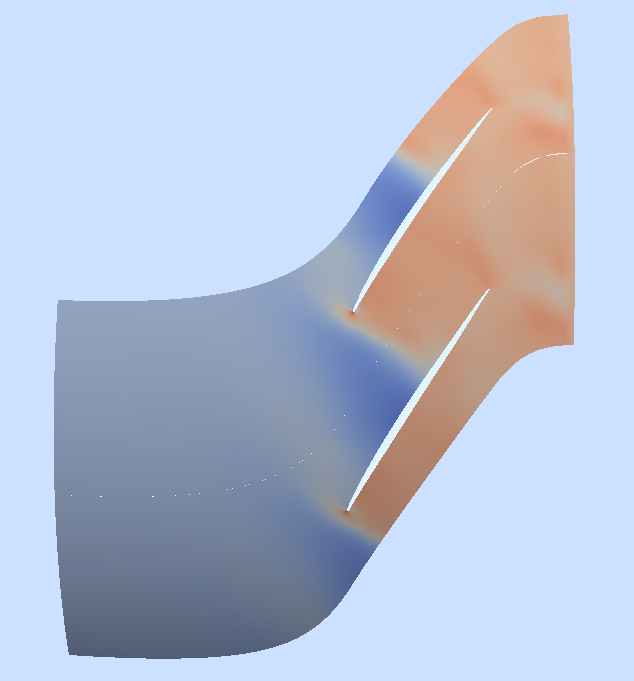}%passageTopviewSolutionEdt.png}
  \caption{Left: Schematic side view of the flow passage. Right: Exemplary pressure distribution at a
    cylindrical cut at 90\% span height between hub and shroud. In order to more clearly see the
    shock forming within the passage, we plot two adjacent passages.
    Red colors imply high pressure, blue colors imply low pressure. }
  \label{flow}
\end{figure}

The flow-domain of the simulation setup and an example output flow field (pressure) at a circumferential
section at a fixed radius is shown in
Figure \ref{flow}.
In order to stabilize the CFD solutions we did not include a clearance
between rotor blade and shroud geometry.
 This is somewhat unrealistic as it qualitatively changes the
tip vortex caused by such a gap and thus overestimates the blade efficiency. But for the current
study it seems a valid simplification which should still provide realistic results except for the
blade geometry in the tip region.

\subsection{Fitness function and optimization algorithms}
\label{secFitness}

The optimization maximizes the isentropic efficiency of the rotor
which is calculated from the OpenFOAM CFD solutions as
\begin{align}
  \label{eqEffDef}
  \eta=\frac{\left(\frac{P_{\m{T, outlet}}}
      {
        P_{\m{T, inlet}}}\right)^\frac{\gamma-1}{\gamma} -1
       }{
         \frac{T_{\m{T, outlet}}}{T_{\m{T, inlet}}} -1
       }
\end{align}
%where $P_\m{T, inlet}$ and $T_\m{T, inlet}$ ($P_\m{T, outlet}$ and $T_\m{T, outlet}$ ) are the mass-flow
%averaged  total pressure and the total temperature at the inlet (outlet) and $\gamma=1.4$ is the heat capacity ratio.
where $P_\m{T}$ and $T_\m{T}$ are the mass-flow
averaged  total pressure and the total temperature at the specified location and $\gamma=1.4$ is the heat capacity ratio.

The fitness function which is minimized during the optimization is given by 
\begin{align}
  f=1-\eta+P
\end{align}
where  the efficiency $\eta$ is averaged over the last 
1000 flow solver iterations and $P$ represents penalty terms which increase (i.e.\ worsens)
the fitness in case the CFD solution 
does not converge or the generated blade geometry is not feasible.  

We employ two different optimization algorithms, the particle-swarm optimization (PSO) \cite{pso}
and the covariance-matrix adaptation strategies (CMA-ES) \cite{cma}. 
Unless otherwise stated, for the CMA-ES we always use a populations size of $\lambda=12$ with $\mu=4$ parents
and an initial step size of $\sigma=0.05$ in relative units where the maximal allowed variation
is normalized to one (i.e., $\sigma=0.05$ amounts to 5\% initial variation).
For the PSO we use 12 particles (which amounts to 12~evaluations per generation)
and the parameters $\omega=0.8$, $\phi_1= 1.7$, and $\phi_2 = 1.4$.
(see Eq.(6) in \cite{pso}).

\section{Results}

\subsection{Comparison between CMA-ES and PSO optimization methods}
\label{secPSO}
In order to evaluate the impact of the optimization algorithm, we compared the results of the two
optimization methods, specifically a ($\mu=4, \lambda=12$) CMA-ES and a PSO with 12 particles.
For the blade deformation, we chose \NHH=9 which
lead to a total of $36$ parameters for the optimizer algorithms.

\begin{figure}[!htb]
  \centering
  \includegraphics[width=0.4\linewidth]{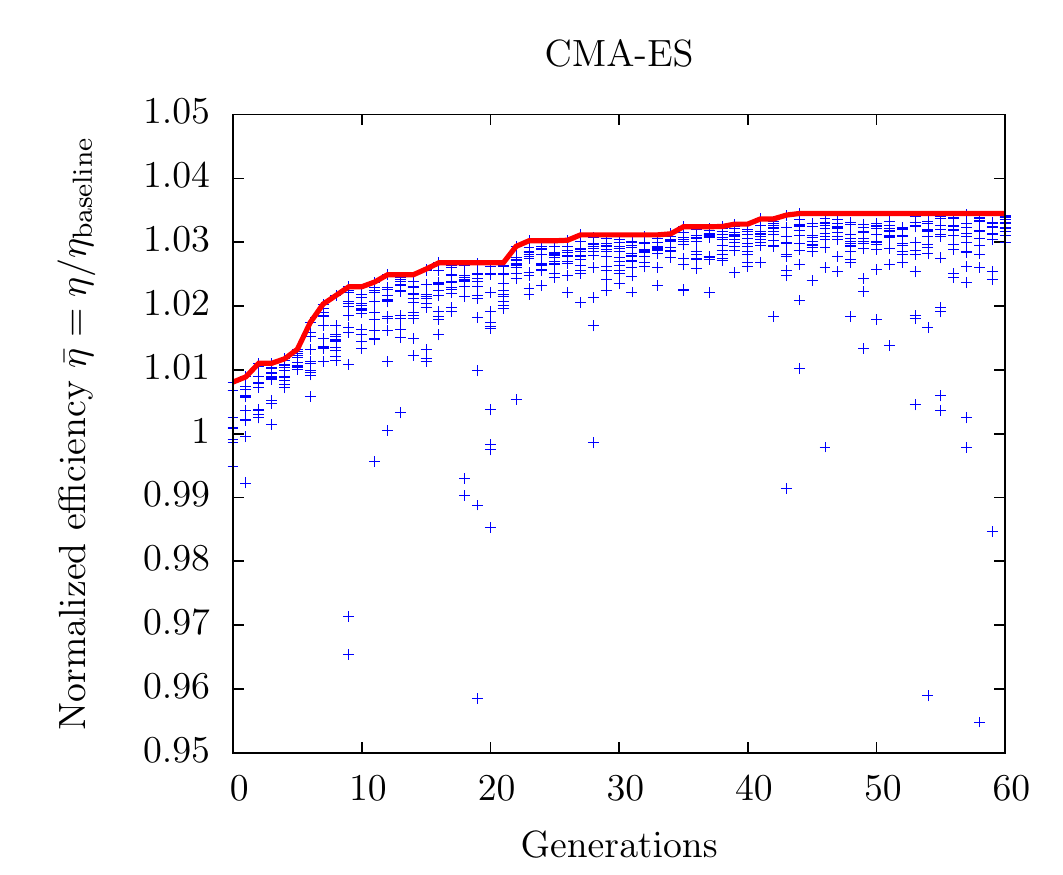}   
  \includegraphics[width=0.4\linewidth]{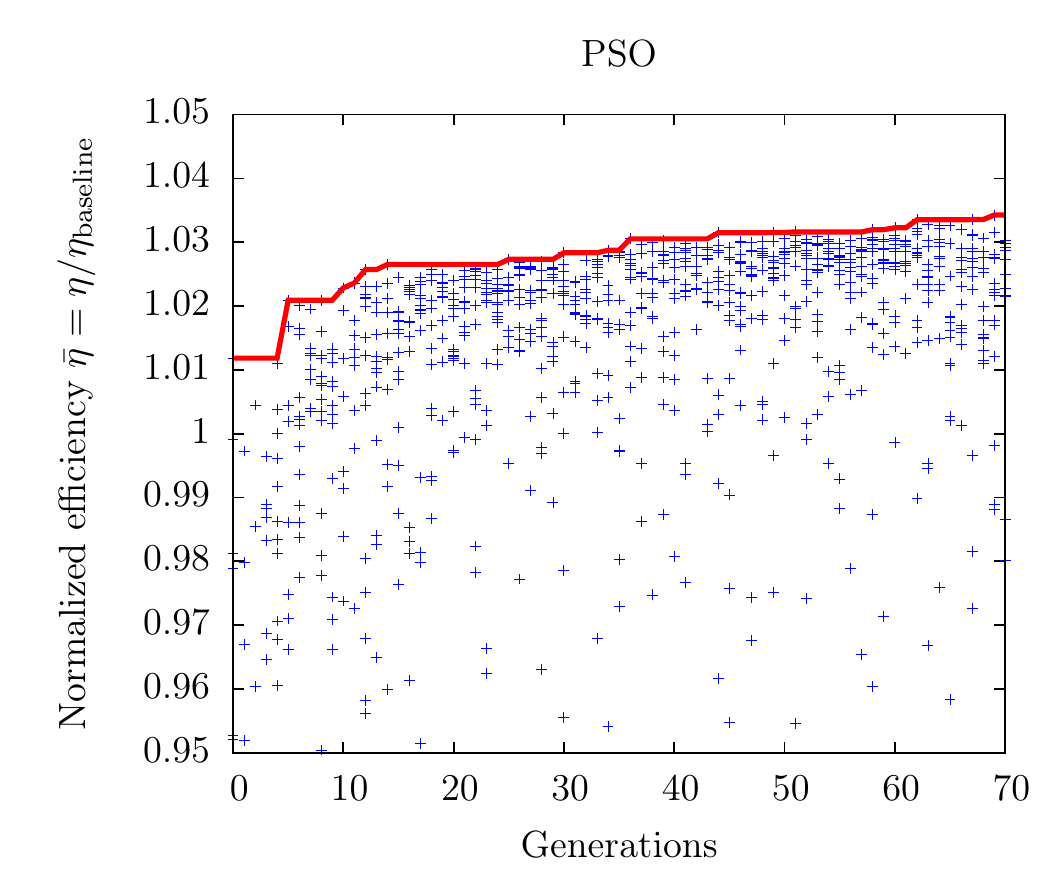} 
  \caption{Normalized efficiency of each evaluated blade geometry as function of
    optimization generations for CMA-ES (left) and PSO (right).
    The red marked solutions just indicate the best solution found so far.}
  \label{CMAvsPSO}
\end{figure}

Figure \ref{CMAvsPSO} shows the progress of the optimization where the efficiency 
of each of the proposed individual geometries is shown as function of the optimization generations.
The efficiency is normalized to the efficiency of the baseline geometry, i.e. $\bar\eta=\eta/\eta_\m{baseline}$.

As can be seen from the Figure \ref{CMAvsPSO}, the variation within the population in each generation
is much larger for the PSO than for the CMA-ES. This is because we used a rather
small initial step size for the CMA-ES, but initialized the particles and velocities for the PSO
randomly within the complete search domain. Thus the PSO performs a more global search, whereas 
the CMA-ES is more local.
However, when looking at the progress of the best solution found up to a certain generation,
both approaches produce very similar results. Also, the fitness values during later stages of the
optimization are approaching very similar results. For example, the best normalized
efficiencies were found to be $\bar\eta_\m{PSO}=~1.0342$ and $\bar\eta_\m{CMA-ES}=1.0345$.

\begin{figure}[!htb]
  \centering
  \includegraphics[width=0.31\linewidth]{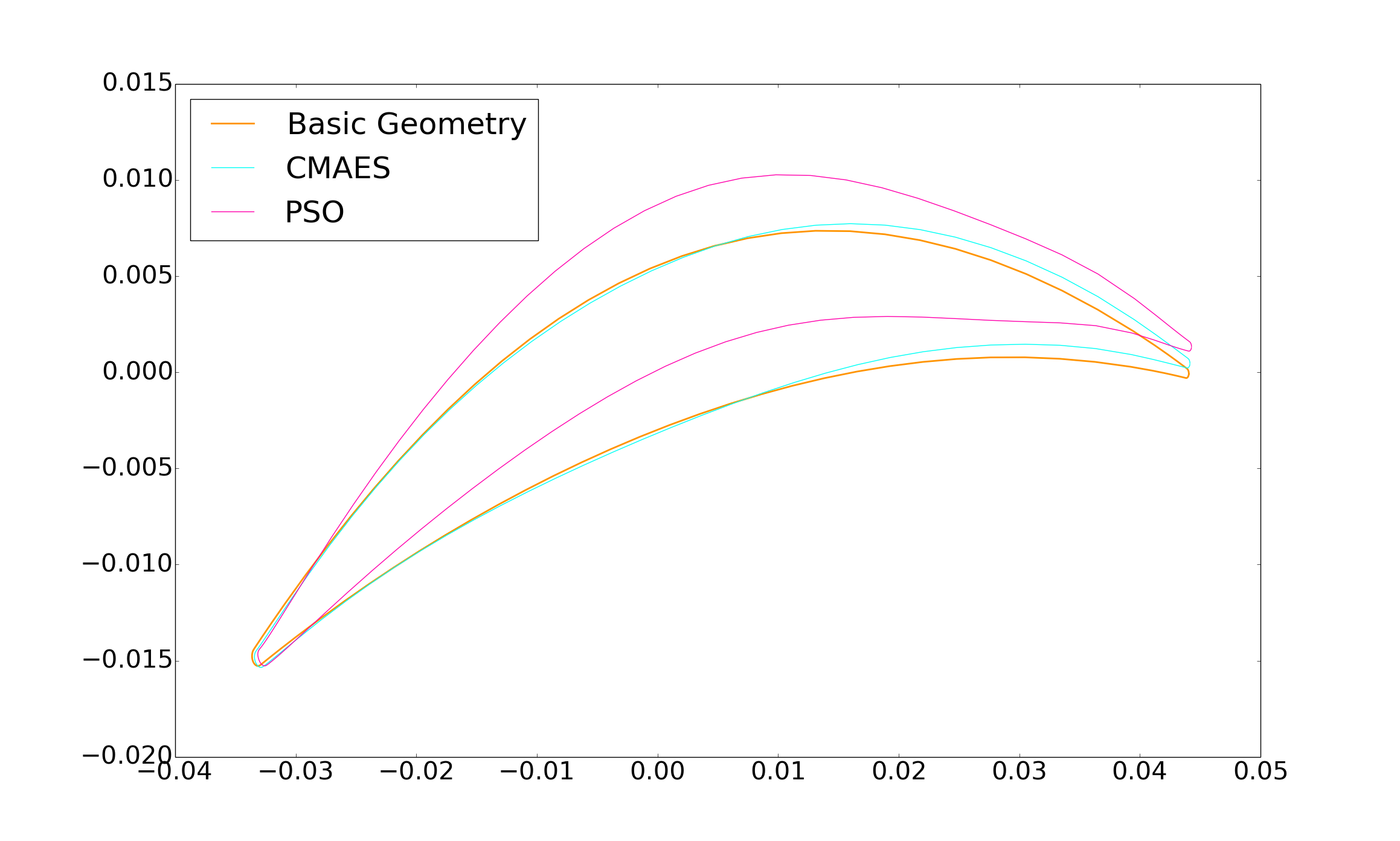} 
  \includegraphics[width=0.31\linewidth]{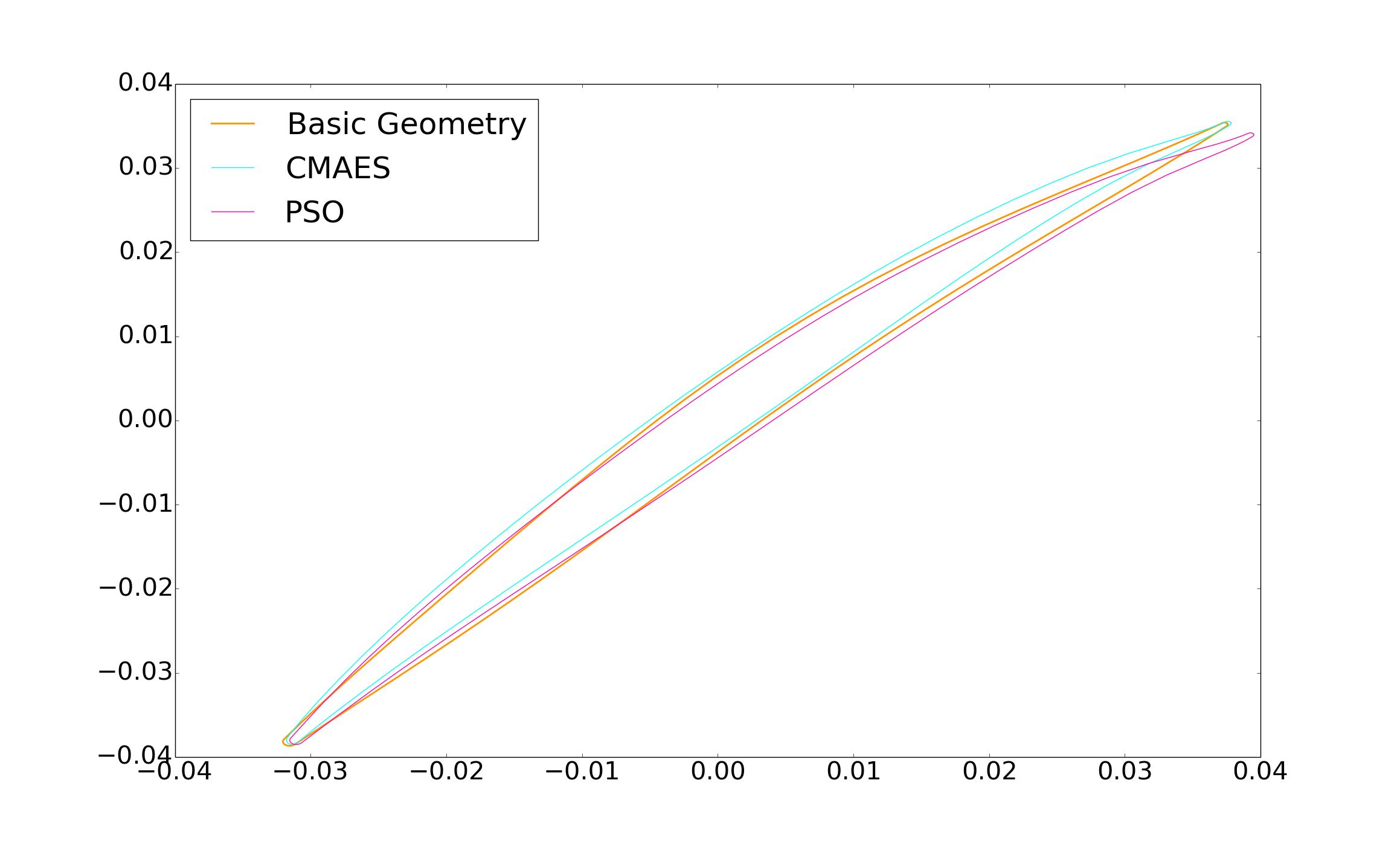} 
  \includegraphics[width=0.31\linewidth]{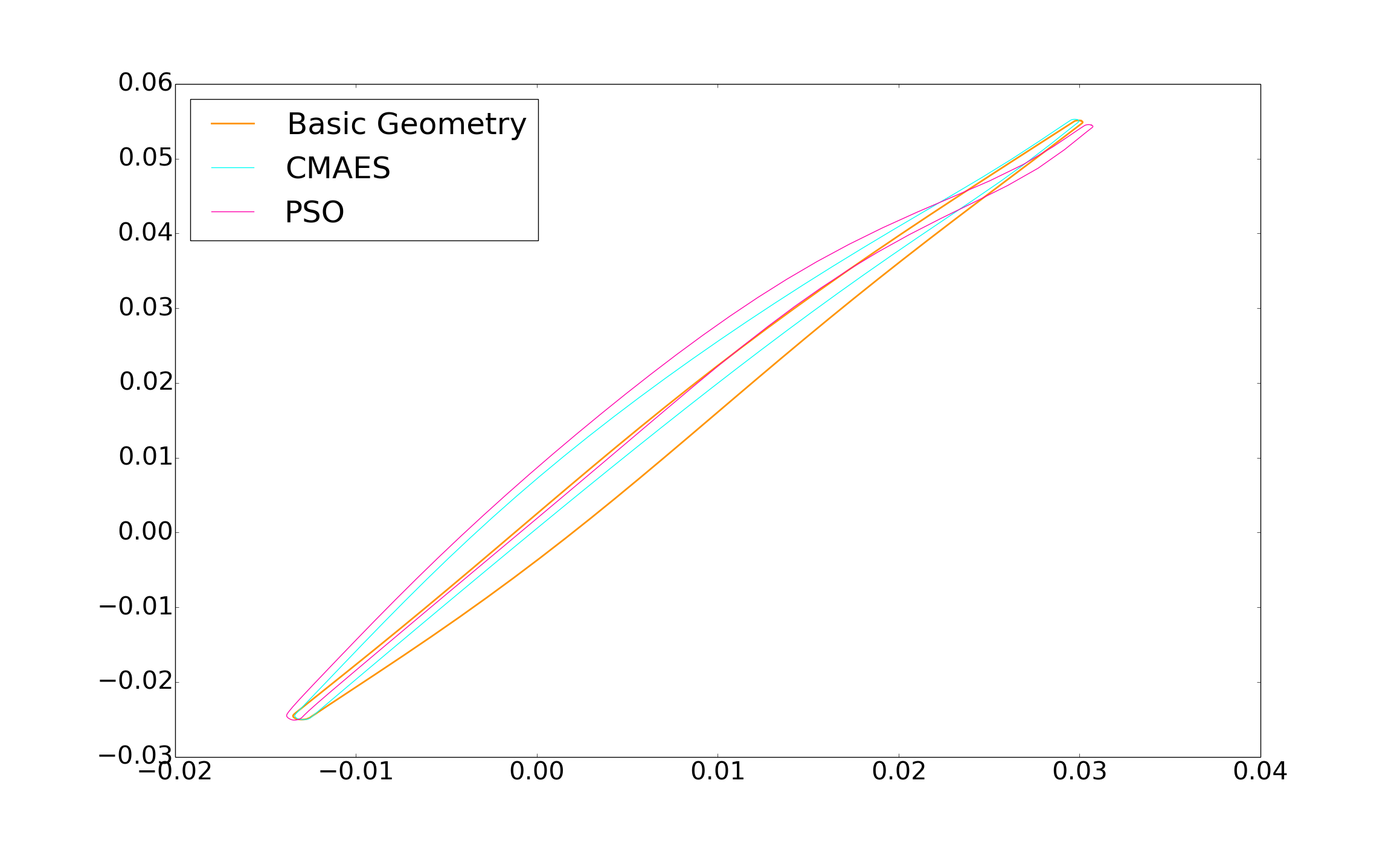}  
  \caption{Comparison of three independently deformed sections  of the final optimized
    designs with the baseline design. The plots are for the hub (left), the mid-span (middle), and the shroud (right) sections.
  }
  \label{CMAvsPSOsects}
\end{figure}

\begin{figure}[!htb]
  \centering     
  \includegraphics[width=0.25\linewidth,height=0.25\linewidth]{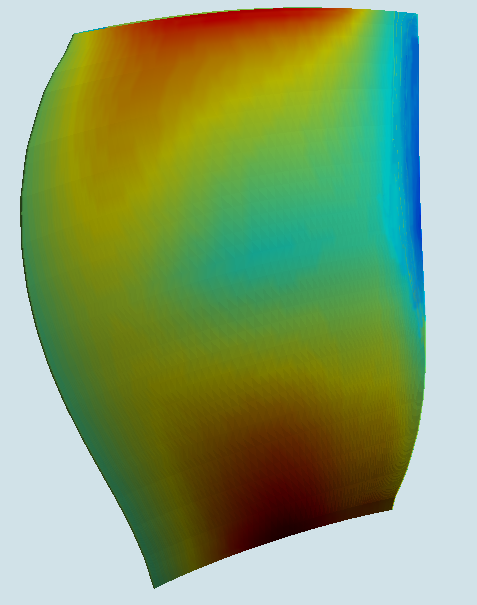} %/6L_crop.png} 
  \hspace*{4mm}
  \includegraphics[width=0.25\linewidth,height=0.25\linewidth]{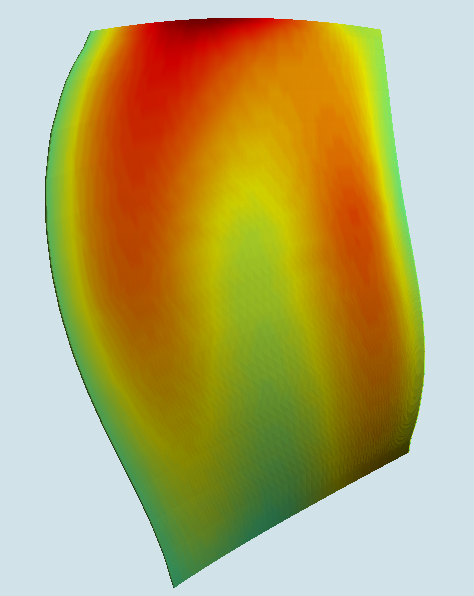} %/6L_crop.png} 
  \caption{The difference between the optimized geometries from PSO (left) and CMA-ES (right) to the baseline design
    where the baseline geometry is shown and the difference is color-coded.
    Green color indicates similar geometries in that region, red indicates the PSO/CMA-ES geometry is
    displaced more into the picture away from the viewer and blue indicates the PSO/CMA-ES geometry
    is displaced toward the viewer as compared to the baseline geometry.}
  \label{CMAvsPSOColorBlade}
\end{figure}

Comparing the actual final optimized geometries, we observe that the two optimization runs
converged to different geometries. As can be seen in Figure \ref{CMAvsPSOsects}
where the three independently deformed sections are shown, all three sections are different
among the baseline, the PSO, and the CMA-ES optimized designs. This leads to substantial differences across
the entire blade, which are nicely visualized
in Figure~\ref{CMAvsPSOColorBlade} where the difference between the geometries is colors-coded.
The PSO and CMA-ES designs show qualitatively different changes to the baseline. For example, the PSO has a
substantially  modified trailing edge region of the fan-rotor whereas the
CMA-ES design is mostly changed in the mid-chord upper region of the blade.

\subsection{Comparison for different random seed initialization}
\label{secSeed}

The stochastic optimization algorithms used in this work need an initialization of
the pseudo-random number generator which is done by specifying a random seed.
The smallest possible change between two optimization
runs is thus given just by using different values for the random seed and leaving everything
else unchanged. 

\begin{figure}[!htb]
  \centering
  \includegraphics[width=0.4\linewidth]{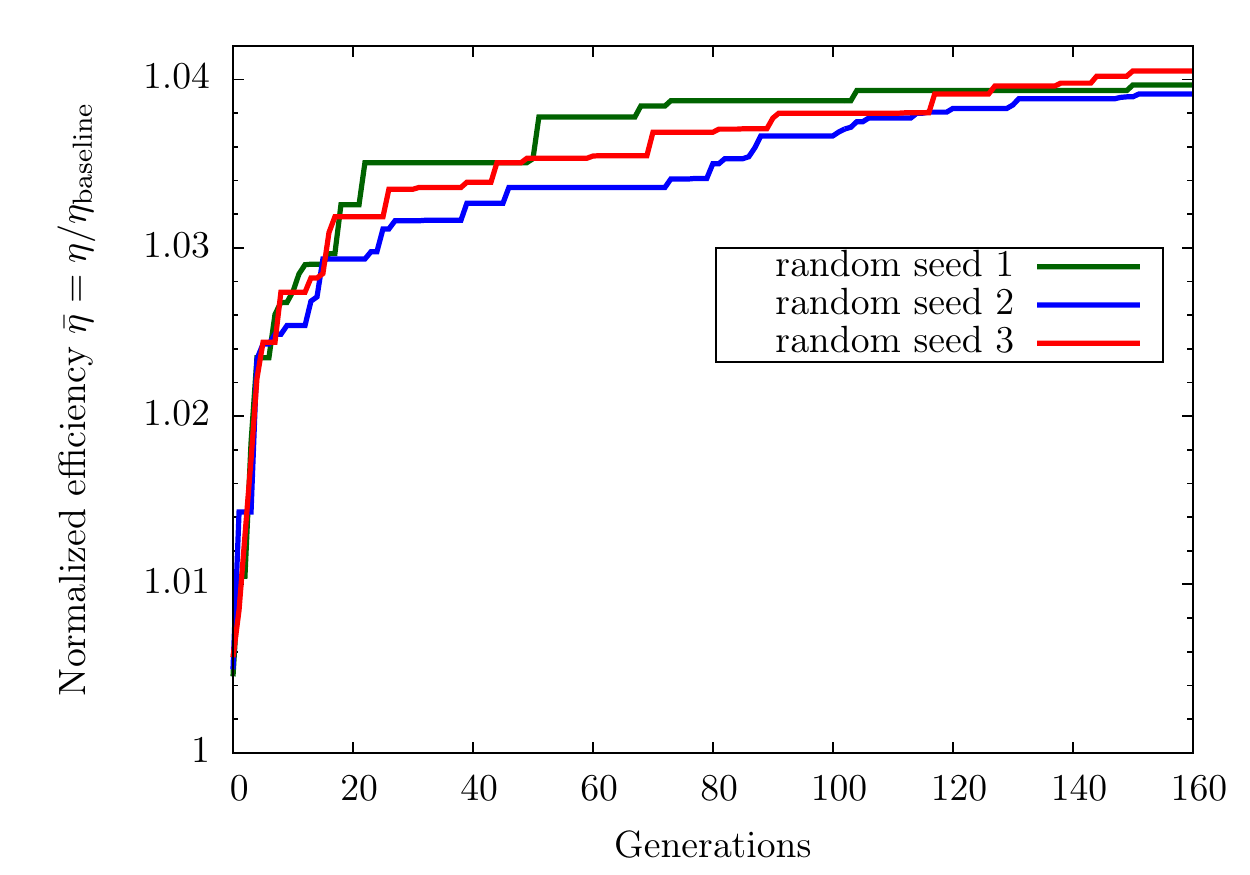}
    % pics/RandomSeedComparison.png 
  \caption{Comparison of best solutions found during the optimization as function of the generation
    of three different CMA-ES optimization runs which just differ in the value
    of the random seed used to initialize the optimization algorithm.}
  \label{ComparisonRandomSeeds}
\end{figure}

We used three different values for the random seed to initialize the CMA-ES optimization run with an otherwise identical setup
with  \NHH=7 which lead to a total of $N_\m{search}=30$ search parameters.
The progress of the best solutions produced by the optimization runs is shown in Figure \ref{ComparisonRandomSeeds}.
It can be seen that the initial progress is quite similar but after some initial generations,
the three runs diverge. During later stages the runs converge again to similar efficiency
values, $\bar\eta_\m{seed1}=1.0397$, $\bar\eta_\m{seed2}=1.0391$, $\bar\eta_\m{seed3}=1.0405$.

\begin{figure}[!htb]
  \centering     
  \includegraphics[width=0.25\linewidth,height=0.25\linewidth]{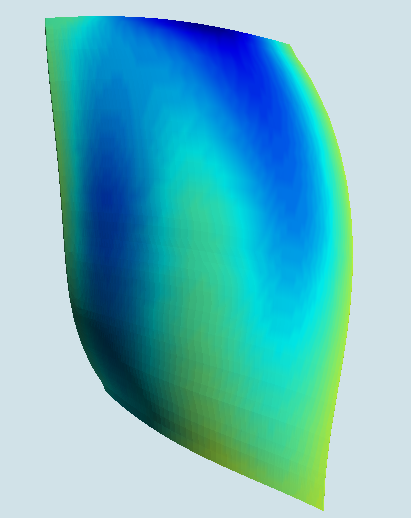} 
  \includegraphics[width=0.25\linewidth,height=0.25\linewidth]{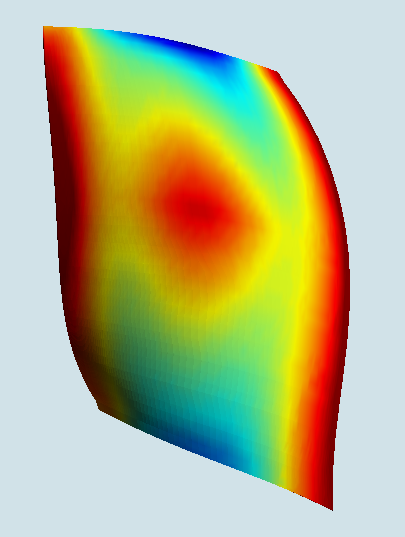}
  \includegraphics[width=0.25\linewidth,height=0.25\linewidth]{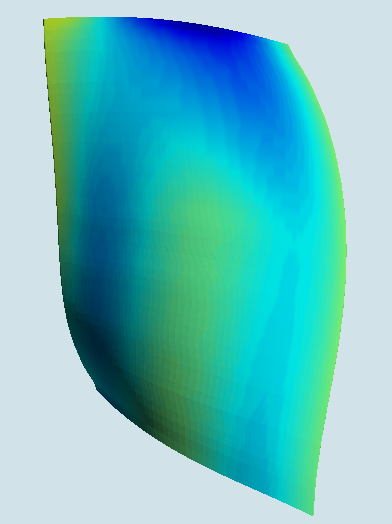}
  \caption{ Color coded differences between the three optimized geometries obtained for three different
    random seeds 1, 2, and 3 (from left tor right) and the baseline geometry. 
    The color coding of the difference is the same as for
    Figure \ref{CMAvsPSOColorBlade}.
  }
  \label{2dplot_p2}
\end{figure}

The comparison of the actual best geometries reveals the same insights as in the previous sections
as the final geometries differ strongly between the runs  (see Figure~\ref{2dplot_p2}).

\subsection{Variation of the number of evaluations per generation}
\label{secCompare}

Another design choice for the optimizer setup is how many evaluations per generation are made. 
Here we compare two CMA-ES runs with ($\mu=4, \lambda=12$) and ($\mu=4, \lambda=24$) and otherwise identical
setup. For the blade deformation, we chose two setups with \NHH=9 and \NHH=12 which
lead to a total of $N_\m{search}=36$ and $N_\m{search}=45$ search parameters, respectively.

The expectation was that the optimization progress would be faster for higher number
of evaluations per generation, if measured in terms of generations since a larger region of the
search space could be sampled in each generation and the information gathered could be used more efficiently.
In practice, with a
usually large resource demand for the CFD simulations, there is  trade-off between
the parallel resources (or licenses) utilized and the overall run-time. This leads to a
limit on the number of evaluations running in parallel.
Therefore, the more relevant measure for optimization performance is in terms of actual evaluations,
i.e.\ number of CFD simulations.

\begin{figure}[!htb]
\centering
  \includegraphics[width=0.4\linewidth]{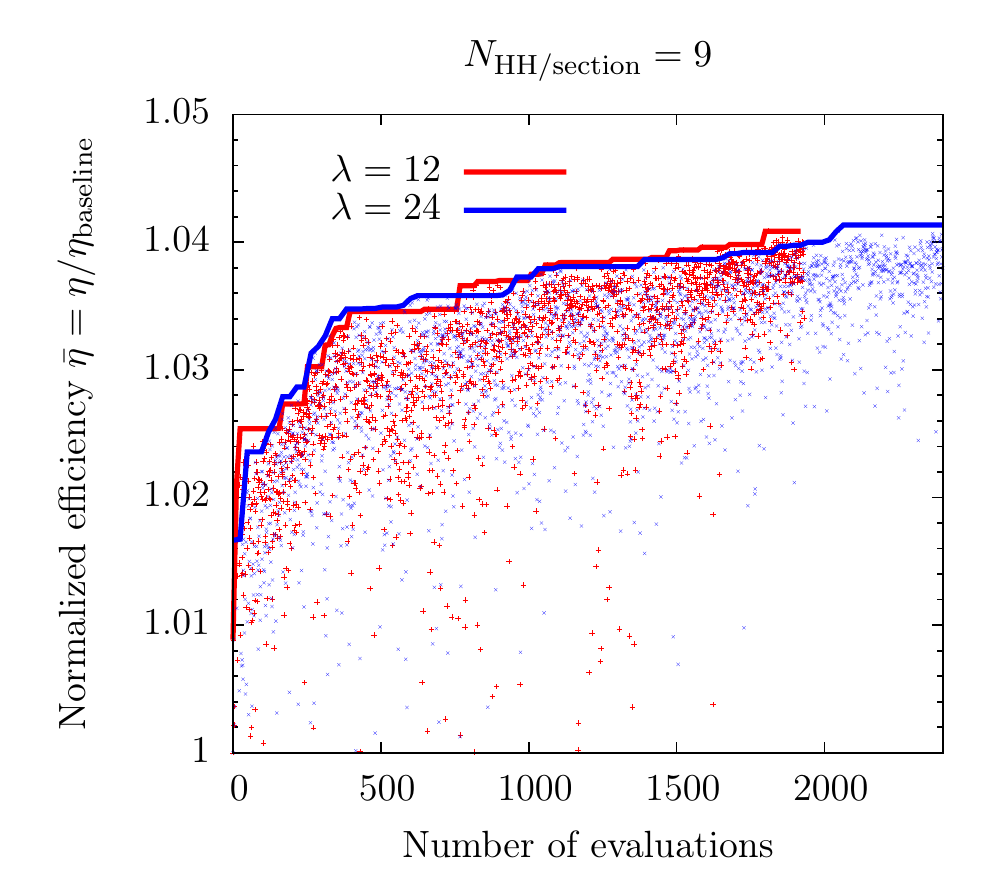} %%pics/HH9_compare.png} 
  \includegraphics[width=0.4\linewidth]{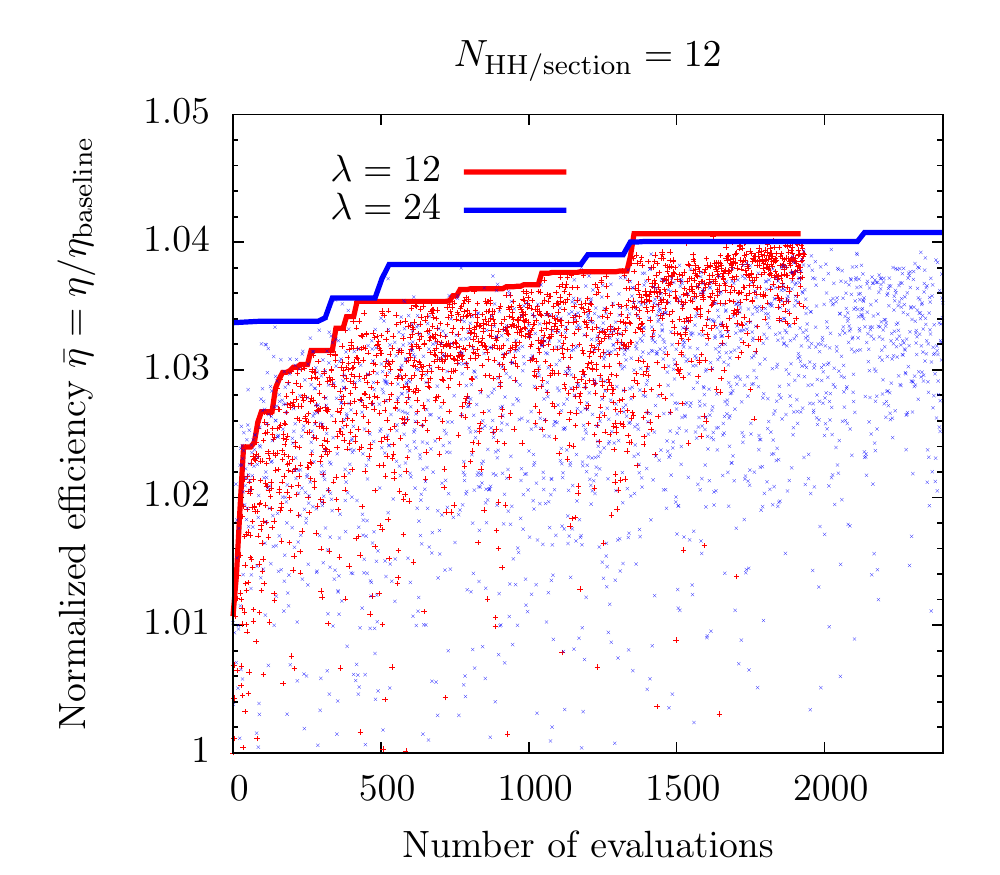} %%pics/HH12_compare.png}    
  \caption{The efficiency as a function of the number of evaluations for  \NHH=9 (left) and \NHH=12 (right) for $\lambda=12$ and $\lambda=24$.}
  \label{diffNumberOfEvals}
\end{figure}

As can be seen in Figure \ref{diffNumberOfEvals}, the optimization progress in terms of actual
evaluations was very similar between the $\lambda=12$ and $\lambda=24$ runs for both setups.
The $\lambda=24$ runs seem to have slightly larger spread in each generation.
It also seems that for larger search dimension, i.e.\ for \NHH=12, a larger population size
is beneficial in the initial phase of the optimization. Both tendencies are somewhat
instructive, but in order to arrive at a really
conclusive evidence, single optimization runs as performed in this work are not sufficient.
Instead, one would need to run each optimization multiple times and compute statistics over the results. 
However, this was not done due to the already high resource demand and
long run-times of the CFD simulations. 

\begin{figure}[!htb]
  \centering
  \includegraphics[width=0.24\linewidth,height=0.25\linewidth]{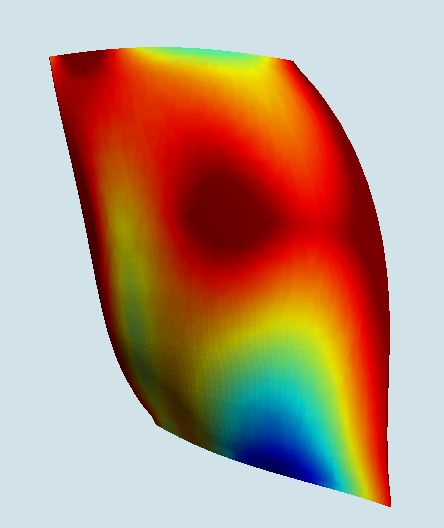}
  \includegraphics[width=0.24\linewidth,height=0.25\linewidth]{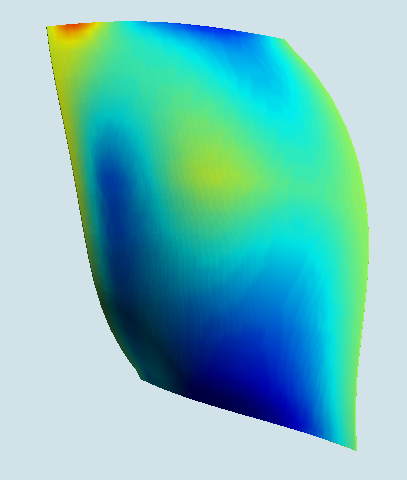}
  \includegraphics[width=0.24\linewidth,height=0.25\linewidth]{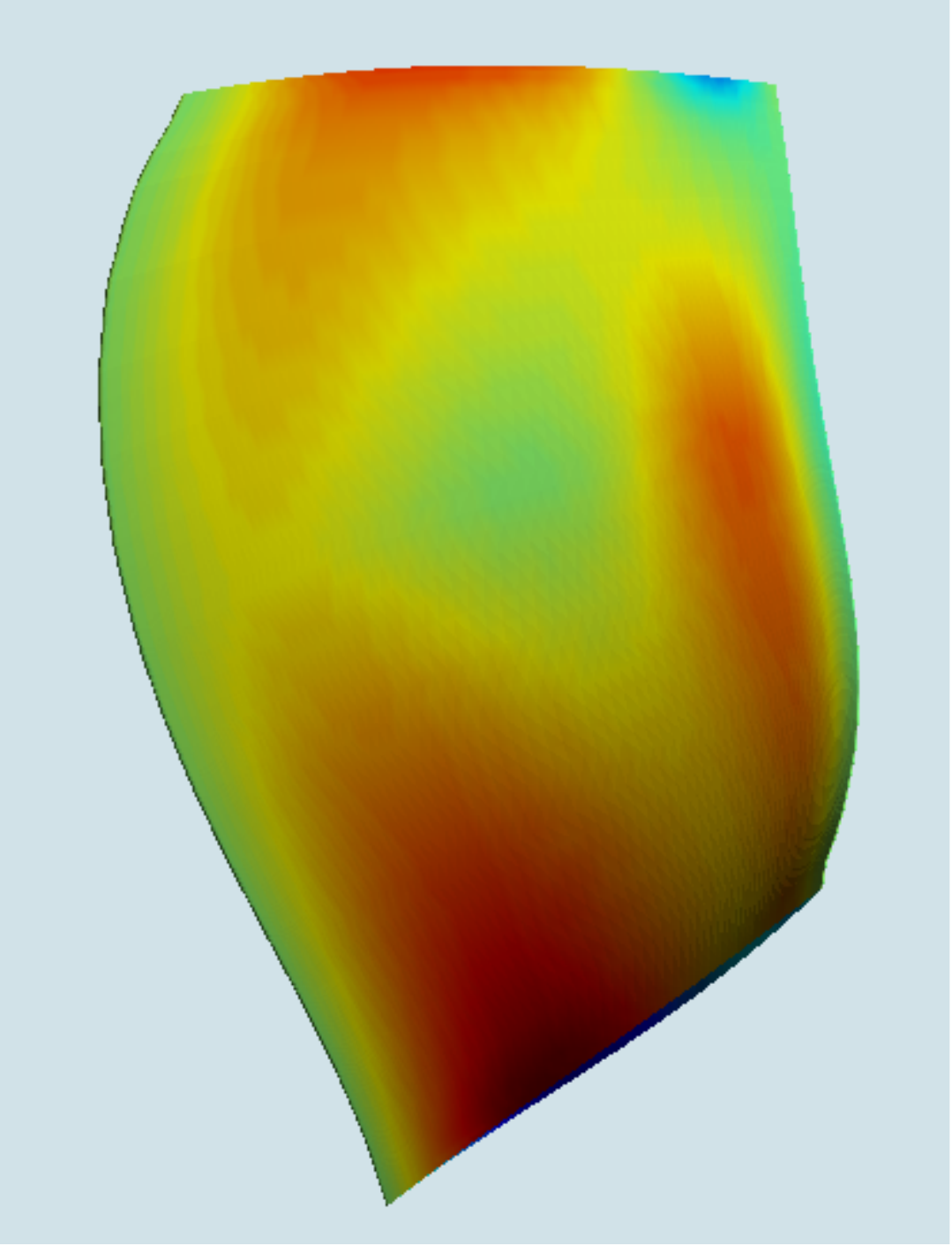}
  \includegraphics[width=0.24\linewidth,height=0.25\linewidth]{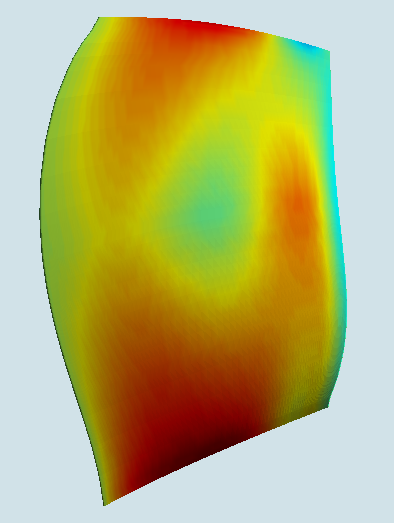}
  \caption{Color coded differences between the baseline geometry and the optimized geometries for
    \NHH=9 and $\lambda=12$,     \NHH=9 and $\lambda=24$, 
    \NHH=12 and $\lambda=12$,   and  \NHH=12 and $\lambda=24$ (from left to right).
    The color coding of the difference is the same as for
    Figure \ref{CMAvsPSOColorBlade}.}
  \label{HHdiffEvals}
\end{figure}

Comparing the achieved efficiency and the actual optimized blade geometries, the same
observations as in the previous sections can be made. Even though the fitness values are comparable,
(\NHH=9: $\bar\eta_\m{\lambda=12}=1.0409$, $\bar\eta_\m{\lambda=24}=1.0414$;
\NHH=12: $\bar\eta_\m{\lambda=12}=1.0407$, $\bar\eta_\m{\lambda=24}=1.0408$),
the resulting geometries differ substantially between the runs over the complete blade for \NHH=9 whereas
the geometries are rather similar for \NHH=12 (see Figure \ref{HHdiffEvals}).

\subsection{Variation of the number of Hicks-Henne shape functions}

A very important choice in the course of blade optimization is the specification of the number of
HH shape functions per section. It is expected that this parameter has a strong influence on
the outcome of the optimization. Choosing a rather small number leads to a very rigid blade
representation where only a limited number of designs can be  realized which might not include
very efficient blades.
On the other hand, a very large number might allow to represent a plethora
of shapes in principle, including very  efficient designs, but the resulting search space
has a very high dimension and thus the optimization problem becomes very hard.
This usually leads to a much slower progress of the optimization and threatens that a sufficiently improved design will not be found within a reasonable
time and the limited resources for the search.

We explored this trade-off by running several optimization runs and varying the number of shape
functions ranging from \NHH=3 up to \NHH=12. This amounted to search dimensions
ranging from  $N_\m{search}=18$ to  $N_\m{search}=45$, and we used the CMA-ES with a populations size of
either $\lambda=12$ or $\lambda=24$.

\begin{figure}[!htb]
\centering
  \includegraphics[width=0.4\linewidth]{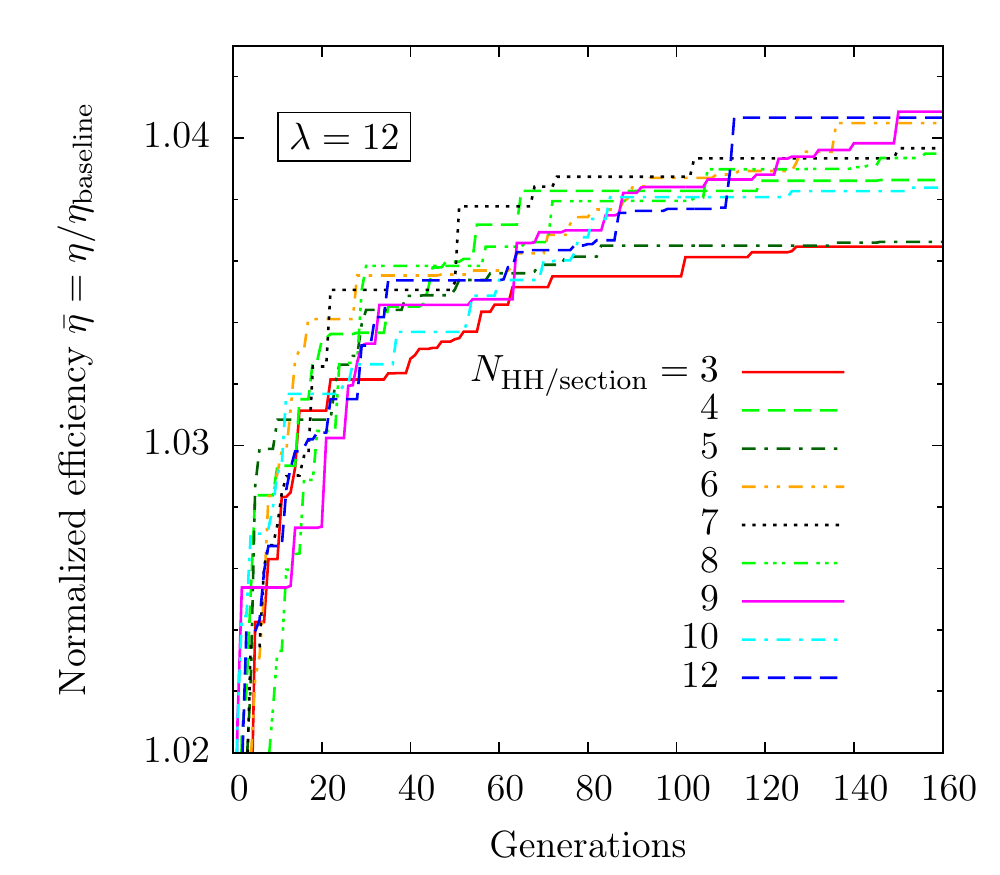} %pics/HH_12evals.png}
  \includegraphics[width=0.4\linewidth]{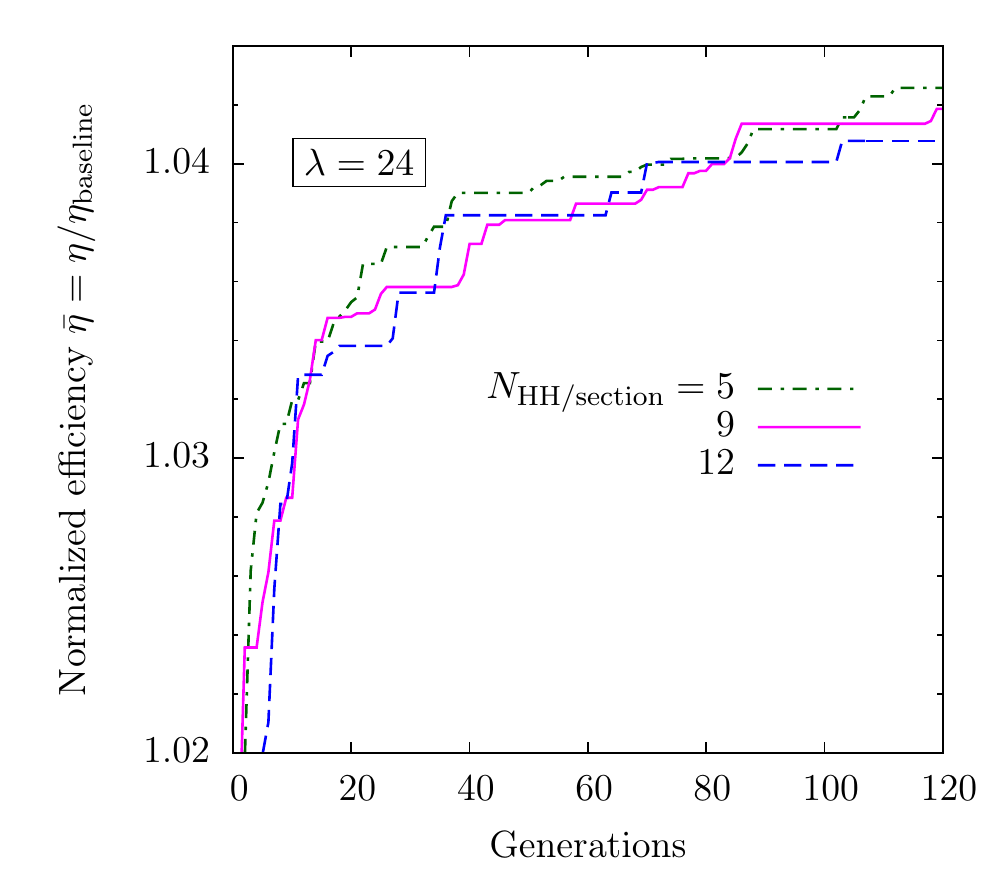} %pics/HH_24evals.png} 
  \caption{The comparison of the best solutions found during the optimization for different number of HH functions for $\lambda=12$ (left) and $\lambda=24$ (right) evaluations per generations.}
  \label{HH12and214}
\end{figure}

The best results as a function of the optimization generations are shown in Figure~\ref{HH12and214}.
First of all it can be observed that there is no clear trend for the achievable efficiency as
function of \NHH. For $\lambda=12$ the smallest number of shape functions per section \NHH=3 indeed
gives the lowest improvement, but for example,  for \NHH=4 a rather large improvement is
realized whereas for \NHH=5 the improvement is as small as for \NHH=3. 
In case of $\lambda=24$, the best efficiency is actually produced with the lowest \NHH=5.

Figure \ref{HH12and214}  suggests that the optimization runs are not yet converged after 120 generations
but instead keep improving with more generations. This is actually expected for 
such population sizes and search dimensions. Therefore, it still could be that there
is a clear relation between the achievable efficiency and \NHH, but for all practical
purposes, with limited computing resources, this is not relevant. The increased
flexibility from a larger number of design parameters is compensated by the increased
difficulty of the search problem.     

The preliminary insight from this is that already with a rather small number of shape functions
per section a considerable improvement in the efficiency can be realized. 
So the choice of \NHH is  not determined by the expected efficiency increase
but rather based on the geometrical aspects implied by the blade representation and the available
optimization run time. 

\begin{figure}[!htb]
  \centering
  \includegraphics[width=0.25\linewidth,height=0.25\linewidth]{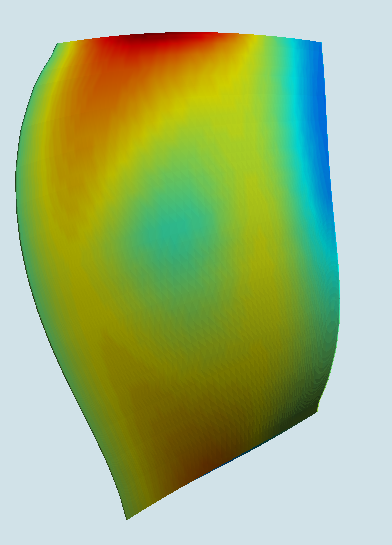}
  \includegraphics[width=0.25\linewidth,height=0.25\linewidth]{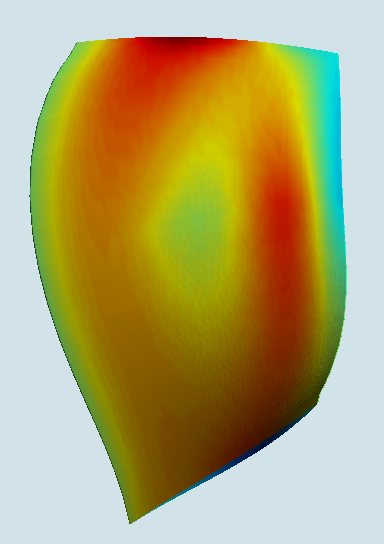} 
  \includegraphics[width=0.25\linewidth,height=0.25\linewidth]{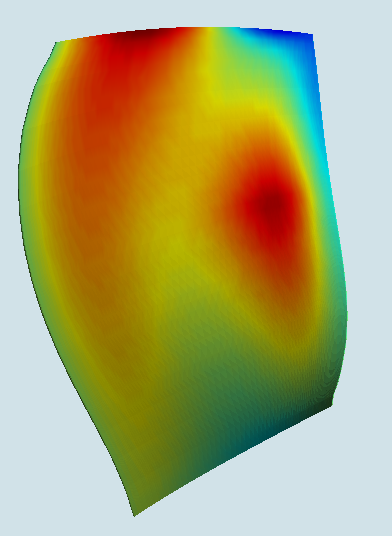}\\[1mm]
  \includegraphics[width=0.25\linewidth,height=0.25\linewidth]{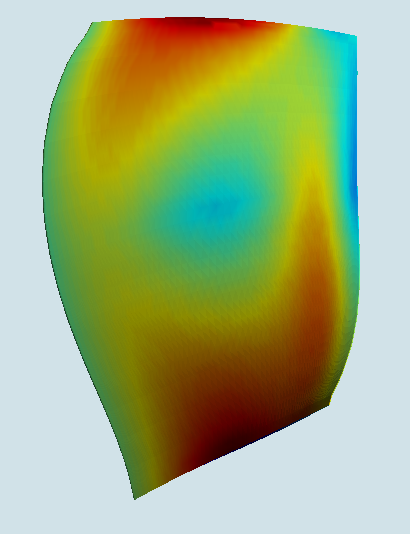} 
  \includegraphics[width=0.25\linewidth,height=0.25\linewidth]{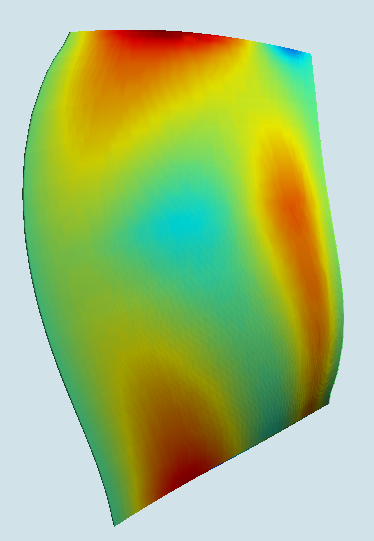} 
  \includegraphics[width=0.25\linewidth,height=0.25\linewidth]{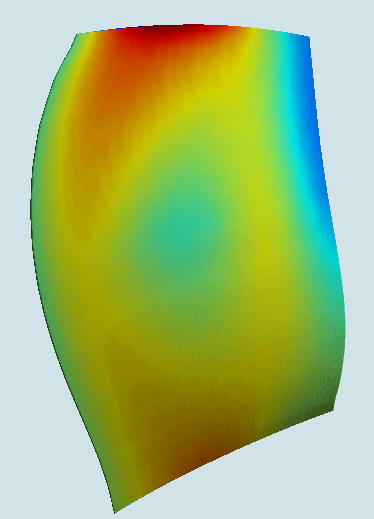} 
  \caption{The comparison of the optimized blade geometries for \NHH=3 (top left),  \NHH=4 (top middle), \NHH=5 (top right),
    \NHH=7 (bottom left), \NHH=10 (bottom middle), \NHH=12 (bottom right) where the difference to the baseline design
    is color coded and displayed on the blade.
    The color-coding is the same as in Figure \ref{CMAvsPSOColorBlade}. 
    \label{HHblades}
  }
\end{figure}

We again observe
that all the resulting geometries are rather different over large parts of the blade which can be seen
in Figure \ref{HHblades}. In particular, 
the shapes are usually  much more curvy and wavy during the optimization for larger values 
of \NHH,
but with increasing generations the shapes seem to be smoothed out again even for larger 
values of \NHH (not shown).

\section{Conclusions}
\label{secConclusion}

We studied multiple shape optimization runs for a realistic turbo-fan
blade geometry setup. In order to explore the fitness landscape, we investigated many setups, varying 
the optimization algorithms, the random seed for the initialization, the population size
and the number of Hicks-Henne shape functions per independently deformable blade section.

The number of Hicks-Henne shape functions does not seem to have a pronounced influence
on the achievable efficiency improvement, at least with the realistic constraint that
only a small number of evaluations (i.e. optimization generations) can be done.
This most likely directly reflects the trade-off between  the increasing 
flexibility with a larger number of design parameters and the increasing
difficulty of the search problem.    
For the practical application this implies that a small number of HH shape functions can already
be sufficient and geometrical aspects for the location and number of shape functions might be 
equally or maybe even more important.

The central finding of this work is that for all the tested variants, comparable improvements in the
rotor efficiency could be  achieved but the actual optimized geometries showed substantial
variation over the complete blade geometry. This leads to the conclusion that the fitness landscape
of such a realistic turbo-fan optimization  is highly multi-modal with many local minima.
Even minor changes (e.g. the random seed for the initialization of the optimization algorithm)
may lead to very different geometries with comparable efficiencies.

This has strong implications for practical blade design optimization approaches. 
On the one hand, the designer can easily generate a set of design variations which all have comparable  
efficiencies  by minor changes to the setup.
On the other hand, this makes surrogate-assisted optimization approaches \cite{qiuKriging2016} 
more difficult. First of all, training a surrogate model on a highly multi-modal function
requires large amounts of data which are most likely not available thereby reducing the model accuracy
drastically. Second, data from finished optimization runs will be dense only around one or a few 
local minima which makes models trained such data not very helpful for optimizations runs which
end up converging to another local minimum.

In the current work we only investigated a single operating condition (fixed mass-flow and rotational speed
at cruise condition) and only the rotor passage. A more realistic scenario should include the downstream 
stator blade which is know to have a strong influence on the fan blade design. Also, additional
  operating conditions with different  mass-flow and rotational speed (e.g. take-off condition) need to be
  incorporated.
The expectation of the authors is that the fitness landscape  of such a more realistic setup will still be
multi-modal, probably to a lesser extent than for the current case, but the general findings of this 
work will still be valid.  Such  investigations are left for future work.

\section*{Acknowledgments}
The authors thank Hisato Tanaka, Kunio Nakakita, Ji\v r\'i \v Simonek, Radek M\'aca and Tom\'a\v s Kr\'atk\'y for valuable
discussion.

\bibliographystyle{amsplain}

\end{document}

%% file: abstract.tex
 Aerodynamic shape optimization has established itself as a valuable tool in the engineering
  design process to achieve highly efficient results. 
  A central aspect for such approaches is the mapping from the design parameters
  which encode the geometry of the shape to be improved to the quality criteria which
  describe its performance. The choices to be made in the setup of the optimization process
  strongly influence this mapping and thus are expected to have a profound influence on the
  achievable result. In this work we explore the influence of such choices on the effects on
  the shape optimization of a turbofan rotor blade as it can be realized within an aircraft engine design
  process. The blade quality is assessed by realistic three dimensional computational fluid dynamics
  (CFD) simulations.
  We investigate  the outcomes of several optimization runs which differ in various configuration options.
  We compare the results from the covariance matrix adaptation evolutionary 
  strategy (CMA-ES) with the outcome of a particle swarm optimization (PSO). 
  We also investigate  the changes induced by a different initialization  of the CMA-ES and 
  by a variation of its population size. 
  A particular focus is put on the variation of the results if we use different number of degrees 
  of freedom for parametrization of the rotor blade geometry. 
  For all such variations,  we  generally find that the achievable improvement of the blade 
  quality is comparable for most settings and thus rather insensitive to the details of the setup.
  On the other hand,  even supposedly minor changes in the settings, such as using
  a different random seed for the initialization of the optimizer algorithm, lead to very
  different shapes. Optimized shapes which show comparable performance usually differ quite
  strongly in their geometries over the complete blade.  
  Our analyses indicate that the fitness landscape for such a
  realistic turbofan rotor blade optimization is highly multi-modal with many local optima,
  where very different shapes show similar performance. 

%%% Local Variables:                                                    
%%% mode: latex                                                         
%%% TeX-master: paperKmecSchmitt.tex                                                       
%%% End: